# Broadband Defects Emission and Enhanced Ligand Raman Scattering in 0D $Cs_3Bi_2I_9$ Colloidal Nanocrystals

Giuseppe M. Paternò[1‡], Nimai Mishra[2,3‡], Alex J. Barker[1], Zhiya Dang[3], Guglielmo Lanzani[1,4], Liberato Manna[3*] and Annamaria Petrozza[1*]

[1]Istituto Italiano di Tecnologia, Center for Nano Science and Technology, Via Pascoli 70/3, 20133 Milano, Italy.

[2]SRM-University AP, Amaravati, Guntur, AP, India, 522502

[3]Istituto Italiano di Tecnologia, Nanochemistry Department, Via Morego 30, 16163 Genova, Italy

[4]Politecnico di Milano, Dipartimento di Fisica, Piazza Leonardo da Vinci 32, 20133 Milano, Italy

**Excitonic 0D and 2D lead-halide perovskites have been recently developed and investigated as new materials for light generation[1]. Here we report broadband (> 1 eV) emission from newly synthesised zero-dimensional (0D) lead-free colloidal $Cs_3Bi_2I_9$ nanocrystals. We investigate the nature of their emissive states as well as the relative dynamics which are currently hotly debated. In particular, we find that the broadband emission is made by the coexistence of emissive excitons and sub-bandgap emissive trap-states. Remarkably, we observe evidence of enhanced Raman scattering from the ligands when attached to the nanocrystals surface, an effect that we preliminary attribute to strong exciton-ligands electronic coupling in these systems.**

* Correspondence should be addressed to Annamaria Petrozza annamaria.petrozza@iit.it, or to Liberato Manna liberato.manna@iit.it.

‡These authors contributed equally to this work.

# Introduction

Lead-halide perovskites have recently emerged as extremely efficient and low-cost solar absorbers [2], as well as bright emitting materials [3] and gain media [4] for applications in light-emitting diodes and lasers. In this context, the introduction of 0D perovskites made of isolated octahedrons and 2D perovskite made of layered nanostructures [1a, 5] has opened new routes to finely tune their optoelectronic properties, by playing with electronic, spatial and dielectric confinement. These confined systems consisting of discrete conductive metal-halide units (in 0D) or layers (2D) embedded in a dielectric matrix are indeed excitonic materials, owing to the relatively large exciton binding energy values (up to 300 meV [1c, 6]). Therefore, they should generally provide narrow and efficient emission. However, their photoluminescence (PL) very often comes along with a sub-gap broadband (FWHM 0.6 - 0.7 eV) emission whose origin is currently debated. Therefore, despite such broadband emission being potentially an appealing feature (e.g. for white light solid state lighting [1a, 5]) as long as there is no control over it, its exploitation remain unreliable. Another issue which can seriously hamper practical application of these promising semiconductors is the toxicity of lead [7]. In this context, bismuth can be a valid alternative, owing to its isoelectronicity with lead and relative low toxicity and high stability. Such bismuth-based systems have attracted a considerable research attention as stable photovoltaic/photodetector materials in recent years [8]. However, also in this case the characterization of the optical properties has been fragmentary and contradicting. Specifically, $Cs_3Bi_2X_9$ is a semiconductor with an indirect band-gap [9]. Polycrystalline thin films with light absorption onset close to 2.4 eV (520 nm) have shown broad emissive bands centred around 1.9 eV (650 nm), thus, with a large Stoke shift and very long lifetimes, of the order of hundreds of nanoseconds. A similar feature has been also found in double perovskite Bi-Ag compounds [10]. In both cases, it has been assigned to band-to-band emission from the virtually not allowed electronic transition. Other reports on single crystalline samples show weak, high energy (around 3.1 eV, thus close to the absorption onset), structured PL at room temperature [11] and 13 K [12]. The authors attribute such emission to excitonic recombination mediated by the lattice phonons, despite no clear and regular energy spacing of the phonon replica has been observed. On the other hand, reports on $MA_3Bi_2X_9$ and $Cs_3Bi_2X_9$ ligands-capped nanocrystals (NCs) either evidence featureless and relatively narrow PL[13] (at 560 nm, FWHM 45 nm) or a dual-spectral PL feature at room temperature with two spikes at 2.14 eV and 2.05 eV, which have been explained by invoking simultaneous indirect and direct band-gap transitions[14].

Here, we report broadband PL emission from zero-dimensional (0D) lead-free colloidal $Cs_3Bi_2I_9$ NCs. By employing ultrafast transient absorption and photoluminescence spectroscopies, we attribute such a broad PL to the coexistence of free excitons and sub-band trap-states emission. Furthermore, the appearance of a well-defined fine structure in the excitonic emission at RT is consistent with a strong interaction between the excitons and ligand vibrations. Finally, we clearly observe the presence of ligands Raman scattering peaks in the emission spectra, a process that is unprecedented for these materials and can be related to the enhancement of ligands Raman modes when coupled to the NCs.

**Results and discussion**

The $Cs_3Bi_2I_9$ NCs were synthesized by a hot injection process. In brief, Cs-Oleate in octadecene (ODE) was injected in the dissolved solution of $BiI_3$ with appropriate amount of surface capping groups in ODE as a solvent. After the injection of the Cs precursor at ∼ 100 °C an immediate colour change from yellow brown to orange-red marked the nucleation of the $Cs_3Bi_2I_9$ NCs. Figure 1a shows representative TEM images of the as synthesized NCs with uniform shape. The XRD patterns of the as-prepared $Cs_3Bi_2I_9$ films, obtained by drop casting NCs dispersions, are shown in Figure 1b. The peaks at 12.5°, 21° and 26° correspond to (011), (110), and (022) crystalline planes according to the calculated XRD pattern of $Cs_3Bi_2I_9$, which showed the typical hexagonal P63/mmc space group. In this structure, $BiI_6$ octahedra share faces to form $[Bi_2I_9]^{3-}$ anions, resulting in a zero-dimensional (0D) molecular salt crystal structure[12]. The azimuthal integration of the selective area electron diffraction (SAED) pattern matched with both hexagonal and monoclinic $Cs_3Bi_2I_9$ phases (see Figure 1c). To obtain further insights over the structure of our NCs, we carried out high-resolution TEM (HRTEM) analysis, shown in Figure 1d, e. It evidenced that the spots in the corresponding fast Fourier transform (FFT) of the NC matched with a monoclinic $Cs_3Bi_2I_9$ phase (ICSD 411633) better than the hexagonal phase. Figure 1f shows the high angle annular dark field (HAADF) – scanning TEM (STEM) image of a group of $Cs_3Bi_2I_9$ NCs, in which the small nanoparticles composed of Bi were induced by electron irradiation. The elemental maps acquired from energy dispersive X-ray spectroscopy (EDS) are reported in Figure 1g and the quantification of the corresponding EDS spectrum revealed that the ratio of Cs:Bi:I is close to 3:2:9, although the NCs are slightly deficient in I, which arises due to the halogen desorption during imaging[15].

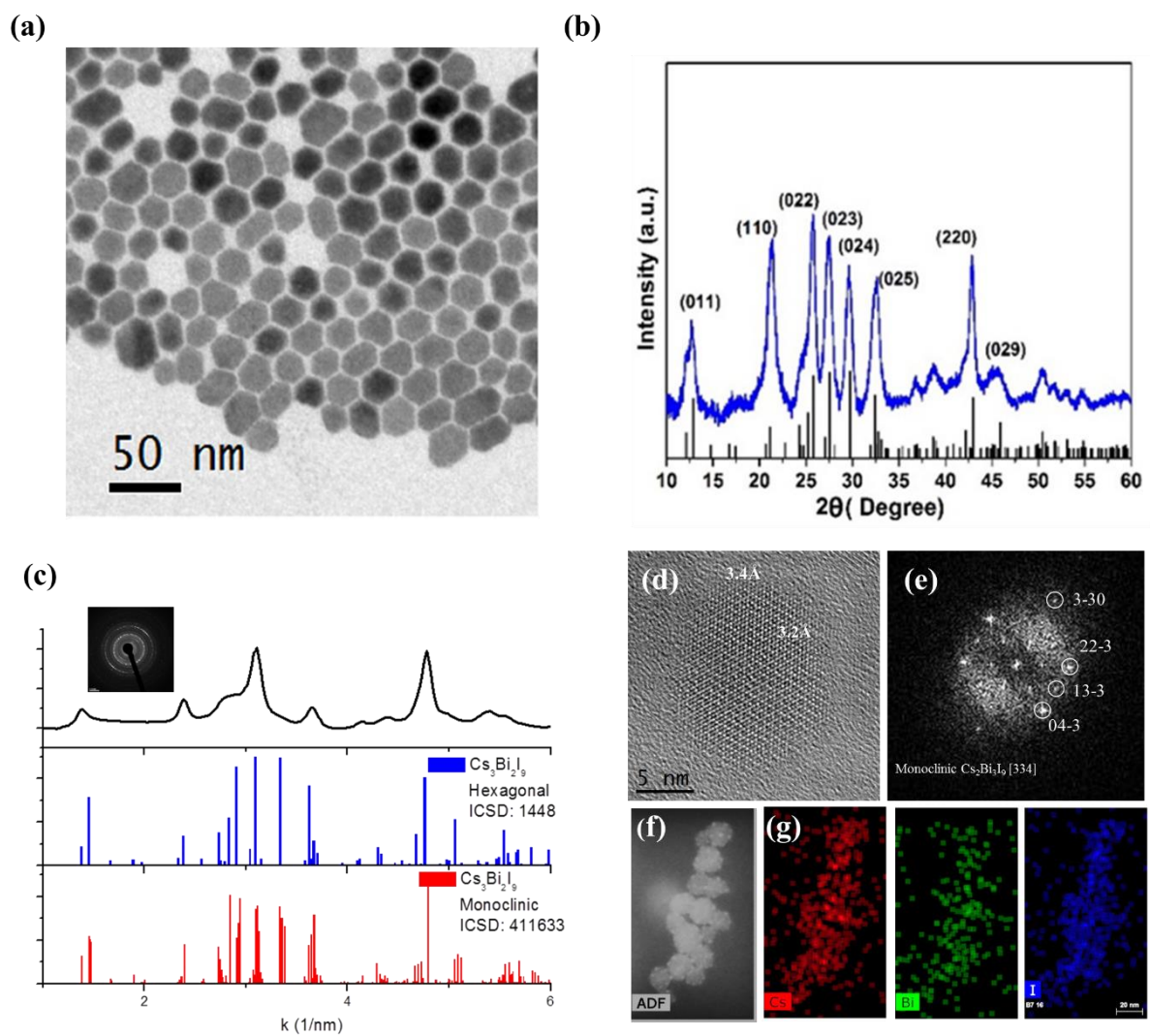

**Figure 1| Structural and compositional characterisation of Cs$_3$Bi$_2$I$_9$ NCs.** (a) Low-resolution transmission electron microscopy (TEM) image of monodispersed Cs$_3$Bi$_2$I$_9$ NCs with 20.359 ± 4.711 nm average diameter. (b) XRD patterns for the NCs. (c) Azimuthal integration of the SAED pattern (inset), in comparison with reference cards of hexagonal and monoclinic Cs$_3$Bi$_2$I$_9$. (d) HRTEM image of a Cs$_3$Bi$_2$I$_9$ NC, and (e) the corresponding fast Fourier transform, whose crystal structure matches with Cs$_3$Bi$_2$I$_9$ with monoclinic phase. (f) HAADF-STEM image of Cs$_3$Bi$_2$I$_9$ NCs and (g) the corresponding EDS elemental maps.

The UV-VIS absorption spectrum of Cs$_3$Bi$_2$I$_9$ NCs in toluene dispersion (Figure 2a) features a relatively sharp peak centred at 2.5 eV and a broad absorption bands extending up to 4 eV containing at least two sub-bands at 3.10 eV and 3.50 eV, which have been ascribed to the electron transitions from the ground $1S_0$ to the excited states of Bi$^{3+}$ in the isolated Bi$_2$I$_9^{3-}$ clusters[16]. We estimated a wide indirect band-gap of 2.75 eV for such material (Tauc Plot exponent $r$ = 3, see Supplementary Figure 1) a value which is in good agreement with those found in previous reports in literature[17].

The band at 2.5 eV is assigned to the localized exciton transition, which results from the presence of isolated $Bi_2I_9^{3-}$ clusters. Such assignment is corroborated by transient absorption (TA) measurements that monitors the change in transmittance of a probe broadband pulse upon photoexcitation with a pump pulse. The transient spectrum, at very early times after phot-excitation, i.e. t = 1 ps (Figure 2b, excitation 3.1 eV) shows a positive sharp band peaking at 2.52 eV and two negative bands, one sharper at higher energy (2.62 eV) and a broader one at 2.23 eV that extends up to 1.75 eV. Such a derivative-like shape of the transient spectrum is the convolution of a photo-bleaching band (PB) with a strong blue shifted PA, which originates from the shift of the excitonic transition experienced by the probe pulse likely due to the reduction of the exciton binding energy induced by multibody interaction. Therefore, it can be taken as a signature of a localized excitonic transition[18]. Beyond 0.5 ns, we note that a positive signal at 1.9 eV kicks in after the decay of the PA in the low-energy part of the spectrum (inset of Figure 2b). We attribute such signal to the PB of inter-gap trap absorption which is populated in about 1 ns.

At room temperature (RT), the PL spectrum of the NCs dispersion in toluene exhibits complex features (Figure 2c). We observe a relatively low Stokes-shifted (0.1 eV) band exhibiting regular oscillations evenly spaced by 0.11 eV and a tail centred at 1.9 eV. Upon cooling we note the clear and dramatic increase of the lower energy tail whose intensity eventually overwhelms the higher energy structured band (see Figure S2 a,b where we show how the oscillations spacing and Stokes shift have been extracted). This marked dependency on the temperature, taken together with the spectral matching with the sub bandgap band at 1.9 eV observed in the TA spectrum, suggest that such emission stems from trap-states[1d] that become more emissive upon decrease of the de-trapping probability at lower temperature[19]. We cannot also exclude the effect of a possible coalescence of the NCs at low temperature which may partially hinder the passivation role of the ligands, highlighting the emission from trap-states. Indeed, the PL spectrum of the drop-cast film is broad and centred at 1.95 eV (FWHM 0.6 eV, Figure S2c) in agreement with previous reports on polycrystalline thin films[8a, 12, 20]. Thus, we can infer that the Stoke shifted low-energy emissive band results from the emissive recombination of trapped carrier, likely on the surface or at the grain boundaries of the NC.

To further assess the different origins of the high and low energy region composing the broad emission spectrum, we carried out power dependency measurements at 78 K (Figure 2d). If we normalise the PL spectra by the excitation power, we observe that the PL centred at 1.9 eV exhibits the typical trend of trap-assisted recombination[19, 21], with an initial increase of the relative PL quantum yield (till 5 mW) due to population of trap-states followed by a dramatic quenching at higher excitation power (80 % decrease at 100 mW). The absence of any hysteretic behaviour of the PL signal (forward and revers scan *vs*. power in Figure S3) confirms that such emission quenching cannot be simply associated to sample degradation.

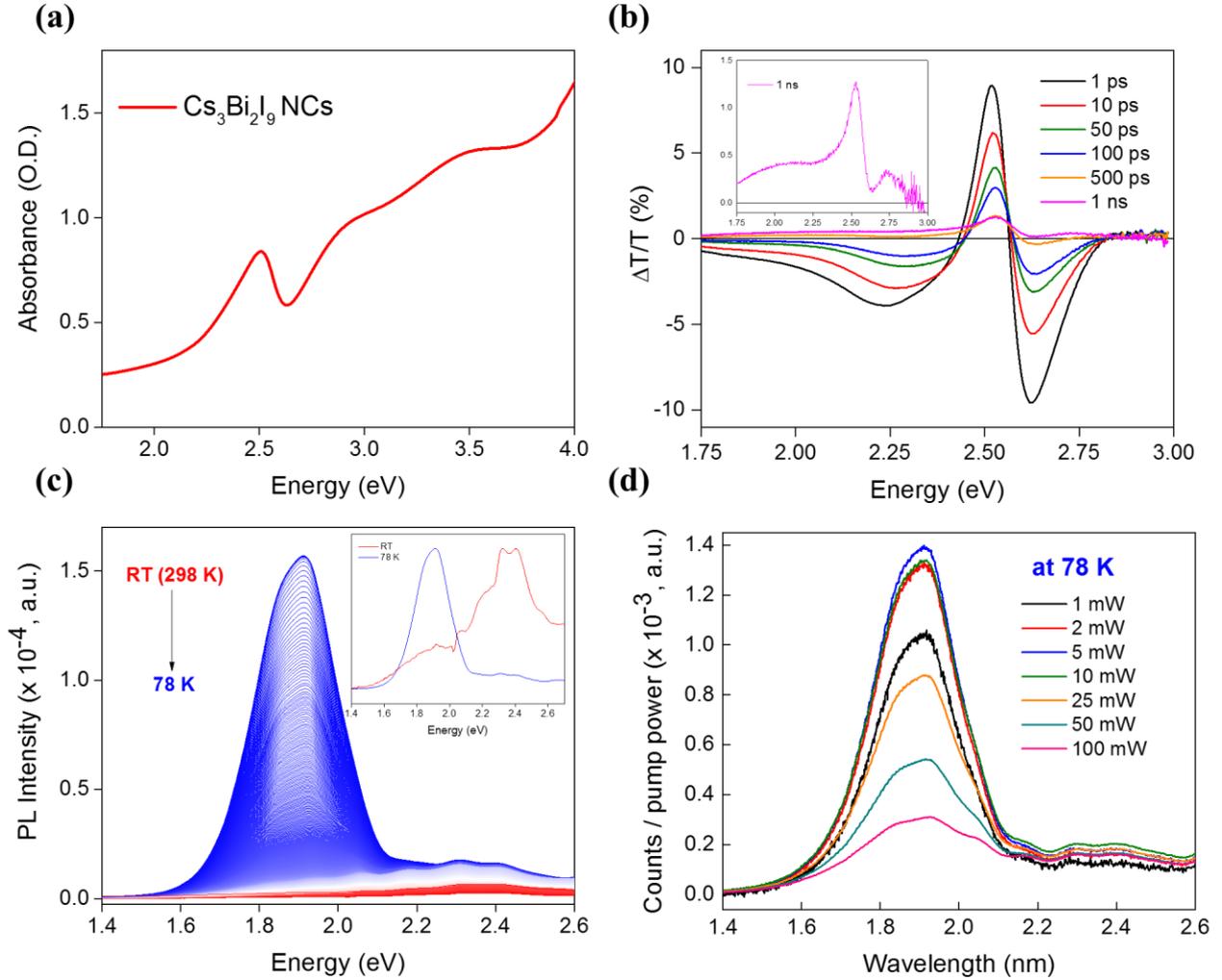

**Figure 2| Spectroscopic characterisation of Cs$_3$Bi$_2$I$_9$ NCs.** (**a**) UV-VIS absorption spectrum of the NCs in toluene dispersion (5 mg/mL). (**b**) Ultrafast transient absorption spectrum obtained by exciting at 3.1 eV of the NCs dispersed in toluene. In the inset we report the TA spectrum at a pump-probe delay of 1 ns, to highlight the appearance of a PB signal at 1.9 eV that we attribute to intra-gap trap-states. (**c**) Photoluminescence spectra recorded upon decrease of the temperature from RT down to 78 K (excitation 3.1 eV, 15 mW). The inset shows the normalised PL at RT and 78 K. (**d**) PL spectra *vs.* excitation power normalised to the excitation power (relative PL quantum efficiency).

After having assessed the presence of different emissive species contributing to the broad PL spectrum, it is worth to further investigate its interesting details. As mentioned above, the band lying in the close proximity of the excitonic absorption, and attributed to free excitons emission, appears to be convoluted with regular oscillations evenly spaced by 0.11 eV which broaden up the otherwise narrow emission from free excitons. The strong tendency of free excitons to interact with lattice phonons and vibrations has been reported previously.[1d, 22] and has been ascribed to the delocalised nature of such excitations that tend to interact with the lattice as a whole, resulting in large spatial overlaps with the phonon wave functions. However, here we measure a constant energy spacing which cannot be attributed to phonon modes belonging to the inorganic moieties, as they normally lie at much higher energies (60-20 meV[1d]). On the other hand, they well match the Raman mode of the oleate ligands (see below).

In Figure 3a we report the PL spectra recorded by exciting at three excitation energies, namely 3.06 eV (in resonance with the band-to-band transition), 2.76 eV (in resonance with the excitonic transition) and 1.96 eV (off-resonance condition). Interestingly, we note a strong and narrow feature on top of the PL band, which is still present even by exciting with an off-resonant energy line. Instead, the excitation of a pure oleate solution in octadecene does not lead to such an effect (Figure S4a). When we plot the emission spectra with the $x$-axis in energy away from the excitation line (excitation – emission, inset of Figure 3a), we see that these three features lie at the same distance from excitation, namely at around 0.36 eV. Surprisingly, we can also discriminate a sharp line incorporated in the broad PL signal of the excitation/emission map (Figure 3b), which emerges at fixed energy from excitation, again at 0.36 eV (red-dotted line). The Raman spectrum of the NCs (Figure 3c) reports strong modes at low frequencies, with two lines at 13 meV and 18 meV that have been attributed to the stretching of each $[BiI_6]^{3-}$ octahedron unit making the $[Bi_2I_9]^{3-}$ cluster and of the Bi-I bond, respectively[1d]. The Raman modes of the oleate ligands lie in the 0.1 eV-0.2 eV and 0.35-0.37 eV regions (for the details on the assignment of Raman modes see ref.[23]). From these data we can confirm two important findings, namely: i) the regular energy spacing on the structured PL of the NCs (0.11 eV) consistently matches with one of the oleate Raman modes ($CH_3$ rocking, chain-end[23]); ii) the strong and narrow peak lying on top of the PL is related to Raman scattering from the ligands (asymmetric stretching of -$CH_2$ at 0.36 eV). Note that we can access mainly to this latter mode because it exhibits the highest shift from excitation among the ligand peaks. The appearance of weak Raman peak at 0.2 eV when exciting off-resonance (orange curve, Figure 1a and inset) that can be attributed to the -C-C stretching mode, might confirm that we can in principle detect all the Raman peaks from simple PL measurements, although most of them are not adequately far from the excitation line. It is also worth adding that we are able to observe Raman scattering by simply employing a fluorimeter equipped with a non-coherent and low-power excitation source (Xenon lamp, 25 mW, see Figure S4b). We can preliminary speculate that the reason of such an enhancement in the ligands-capped $Cs_3Bi_2I_9$ NCs may be contextualised within the surface enhanced Raman scattering (SERS) theory. This would imply that the photo-excited excitons must strongly couple to the ligand vibrations by building up a stable population which is able create a local field effect. The Franck-Condon-like modulation of the PL emission by ligands vibrations further substantiate the conjecture of hybrid coupling between excitons and ligands. The actual exciton-ligand interaction mechanism is however still unclear, although we may tentatively speculate that it could involve either mediated-charge trapping from the excitonic state to the surface-ligand bonding or a complete charge transfer into a ligand moiety[24]. Further experiments and calculations are needed to elucidate these findings.

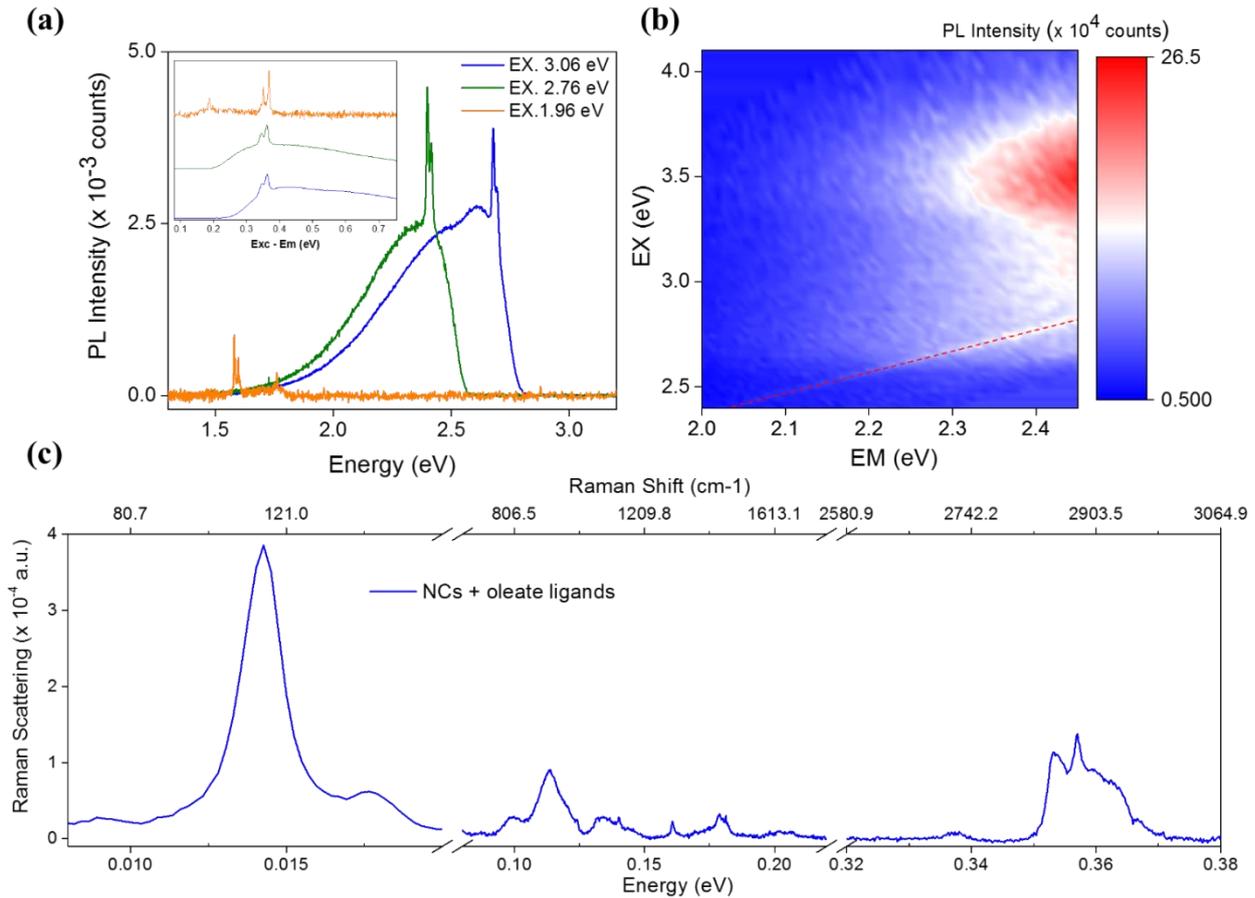

**Figure 3| Ligand modes Raman enhancement in Cs₃Bi₂I₉ NCs.** (**a**) PL spectra take at three excitation lines (3.06 eV, 2.76 eV, 1.96 eV, 15 mW) in toluene dispersion. The inset shows the spectra as a function of excitation – emission energy. Note that for these measurements we dilute the dispersion 10 times from the original concentration (0.5 mg/mL), to suppress self-absorption from the NCs in the Raman scattering region. (**b**) Excitation-emission map taken with a fluorimeter and exciting with a xenon lamp (25 µW). (**c**) Raman spectra taken with a microscope (50 x objective, 10 % power) on NCs drop-cast films.

## Conclusions

In conclusion, we have shown that newly synthesised ligands-capped $Cs_3Bi_2I_9$ NCs can emit broadband light, a process that can be explained in term of concerted emission from free-excitons (low Stokes-shifted PL) and traps-mediated recombination (high Stokes shift PL). The presence of a regular fine structure in the free-exciton PL already at RT suggests that the free exciton population can interact effectively with the ligand vibrations. This in turns may indicate a delocalization of the electronic wavefunction into the ligands shell, possibly improving inter-crystal interaction and transport. We report preliminary evidence of Raman enhancement of the ligand modes when coupled to the NCs, a process that has never been reported for these materials, to the best of our knowledge. Such interesting properties can be extremely appealing for a new class of low-cost materials suitable for solid-state lightning and sensing applications.

## Materials and Methods

*Materials.* All reagents were used without any purification: $BiI_3$ (bismuth triiodide, 99%, Alfa Aesar), $Cs_2CO_3$ (cesium carbonate, 99%, Sigma Aldrich), OA (oleic acid, 90%, Sigma Aldrich), ODE (1-octadecene, 90%, Sigma Aldrich), OLA (oleylamine, 90%, Sigma Aldrich), ODPA (n-octadecylphosphonic acid, 97%, Sterm), and anhydrous toluene (99.98%, Sigma Aldrich).

*Synthesis of $Cs_3Bi_2I_9$ Nanocrystals.* First, Cs-oleate stock solution was prepared in the following way. A mixture of 1.5 gm of $Cs_2CO_3$, 1.5 mL of OA, and 15 mL of ODE were degassed at 100 °C for 2 hr, then the solution was heated at 150°C for 1 hr to make sure the Cs-Oleate formation is complete. In a typical process, the $Cs_3Bi_2I_9$ nanocrystals (NCs) were synthesized via a previously reported modified method of $CsPbBr_3$ NCs synthesis. Briefly, in a 50 mL three-neck round-bottom flask, 0.2 mmol $BiI_3$, 1.2 mmol OLA, 1.2 mmol OA, 0.3 mmol ODPA and 5 mL of 1-ODE were loaded and degassed at 70 °C for about 1 h. The solution was then heated to 100 °C for ∼10−15 min to yield a clear yellow-orange solution. Then 1 ml of Cs-oleate stock solution injected on the yellow-orange solution and cool down immediately. The as-synthesized nanocrystals were collected via precipitated out from the growth solution by centrifuging at 6000 rpm. The red colour precipitate at the bottom was collected after dissolving them in toluene and the supernatant was discarded.

*XRD Characterization.* X-ray Diffraction (XRD) data was obtained with a diffractometer (X-ray diffraction - PANalytical XPert PRO) using Cu−Kα radiation (λ = 1.540598 Å) in the range of 10−80°. Samples were prepared on a clean silicon wafer by placing drops of concentrated nanoparticles in toluene on the silicon surface and dried under vacuum.

*Methods for HRTEM analysis.* Samples were prepared by dropping dilute solutions of nanocrystals onto ultrathin carbon/holey carbon coated 400 mesh copper grids for high Resolution TEM (HRTEM), selective area electron diffraction (SAED), performed with a JEOL JEM-2200FS microscope equipped with a 200 kV field emission gun, a CEOS spherical aberration corrector for the objective lens and an in-column image filter (Ω-type). The composition was determined by energy dispersive X-ray spectroscopy (EDS) EDS analysis performed on the same microscope in high angle annular dark field scanning TEM (HAADF-STEM) mode with a Bruker Quantax 400 system with a 60 $mm^2$ XFlash 6T silicon drift detector (SDD).

*UV-VIS and PL measurements.* For the UV-VIS absorption measurements, we used a Perkin Elmer Lambda 1050 spectrophotometer, equipped with deuterium (180-320 nm) and tungsten (320-3300 nm) lamps and three detectors (photomultiplier 180-860 nm, InGaAs 860-1300 nm and PbS 1300-3300 nm). All the absorption spectra were corrected for the reference spectra taken at 100% transmission (without the sample) at 0% transmission (with an internal attenuator), and for the background spectrum (toluene only). For the PL measurements, we excited with continuous wave (CW) diode lasers (OXXIUS) at 405 nm, 450 nm and 633 nm. The PL was collected perpendicular to the excitation line and focused into a fibre coupled to a spectrometer (Ocean Optics Maya Pro 2000). For relative PLQY, the integrated PL was measured at varying excitation intensities and plotted as: Relative PLQY = $PL/I_{pump}$. To measure the Raman peak incorporated into the PL band (Figure 3a) we diluted the toluene dispersion 10 times (0,5 mg/mL), to decrease self-absorption phenomena in the Raman scattering region (Raman shift = 0.36 eV) and appreciate better such an effect. The PL measurements with the fluorimeter were taken with a Horiba Nanolog Fluorimeter, equipped with a Xenon lamp, two monochromators and two detectors (photomultiplier and InGaAs).

*Raman measurements.* For the Raman measurements we used were recorded in the range of 50-3200 cm$^{-1}$ with a micro Raman confocal microscope (inVia Raman Microscope Renishaw, 10 x objective, 532 nm excitation wavelength). All measurements were carried out in air.

## Acknowledgements

This is the pre-peer reviewed version of the following article: *Broadband Defects Emission and Enhanced Ligand Raman Scattering in 0D Cs$_3$Bi$_2$I$_9$ Colloidal Nanocrystals*, which has been published in final form at 10.1002/adfm.201805299. This article may be used for non-commercial purposes in accordance with Wiley Terms and Conditions for Use of Self-Archived Versions.

We thank the financial support from the EU Horizon 2020 Research and Innovation Programme under Grant Agreement N. 643238 (SYNCHRONICS).

## References

[1] a) D. Cortecchia, S. Neutzner, A. R. Srimath Kandada, E. Mosconi, D. Meggiolaro, F. De Angelis, C. Soci, A. Petrozza, *J. Am. Chem. Soc.* **2017**, 139, 39; b) C. Zhou, H. Lin, Y. Tian, Z. Yuan, R. Clark, B. Chen, L. J. van de Burgt, J. C. Wang, Y. Zhou, K. Hanson, Q. J. Meisner, J. Neu, T. Besara, T. Siegrist, E. Lambers, P. Djurovich, B. Ma, *Chem Sci* **2018**, 9, 586; c) M. D. Smith, H. I. Karunadasa, *Acc. Chem. Res.* **2018**, 51, 619; d) A. Nilă, M. Baibarac, A. Matea, R. Mitran, I. Baltog, *physica status solidi (b)* **2017**, 254, 1552805.e) Jawaher M Almutlaq, Jun Yin, Omar F Mohammed, Osman M Bakr, J. Phys. Chem. Lett. 2018, 9, 14, 4131

[2] a) M. M. Lee, J. Teuscher, T. Miyasaka, T. N. Murakami, H. J. Snaith, *Science* **2012**, 338, 643; b) S. D. Stranks, G. E. Eperon, G. Grancini, C. Menelaou, M. J. Alcocer, T. Leijtens, L. M. Herz, A. Petrozza, H. J. Snaith, *Science* **2013**, 342, 341.

[3] Z. K. Tan, R. S. Moghaddam, M. L. Lai, P. Docampo, R. Higler, F. Deschler, M. Price, A. Sadhanala, L. M. Pazos, D. Credgington, F. Hanusch, T. Bein, H. J. Snaith, R. H. Friend, *Nat Nanotechnol* **2014**, 9, 687.

[4] H. Zhu, Y. Fu, F. Meng, X. Wu, Z. Gong, Q. Ding, M. V. Gustafsson, M. T. Trinh, S. Jin, X. Y. Zhu, *Nat Mater* **2015**, 14, 636.

[5] E. R. Dohner, A. Jaffe, L. R. Bradshaw, H. I. Karunadasa, *J. Am. Chem. Soc.* **2014**, 136, 13154.

[6]  H. Takagi, H. Kunugita, K. Ema, *PhRvB* **2013**, 87, 125421.
[7]  D. Fabini, *J Phys Chem Lett* **2015**, 6, 3546.
[8]  a) B. W. Park, B. Philippe, X. Zhang, H. Rensmo, G. Boschloo, E. M. Johansson, *Adv. Mater.* **2015**, 27, 6806; b) C. Ran, Z. Wu, J. Xi, F. Yuan, H. Dong, T. Lei, X. He, X. Hou, *The journal of physical chemistry letters* **2017**, 8, 394.
[9]  K.-H. Hong, J. Kim, L. Debbichi, H. Kim, S. H. Im, **2016**.
[10] a) B. A. Connor, L. Leppert, M. D. Smith, J. B. Neaton, H. I. Karunadasa, *J. Am. Chem. Soc.* **2018**, 140, 5235; b) S. E. Creutz, E. N. Crites, M. C. De Siena, D. R. Gamelin, *Nano Lett.* **2018**, 18, 1118; c) Y. Bekenstein, J. C. Dahl, J. Huang, W. T. Osowiecki, J. K. Swabeck, E. M. Chan, P. Yang, A. P. Alivisatos, *Nano Lett.* **2018**, 18, 3502.
[11] K. K. Bass, L. Estergreen, C. N. Savory, J. Buckeridge, D. O. Scanlon, P. I. Djurovich, S. E. Bradforth, M. E. Thompson, B. C. Melot, *Inorg. Chem.* **2017**, 56, 42.
[12] K. M. McCall, C. C. Stoumpos, S. S. Kostina, M. G. Kanatzidis, B. W. Wessels, *Chem. Mater.* **2017**, 29, 4129.
[13] B. Yang, J. Chen, F. Hong, X. Mao, K. Zheng, S. Yang, Y. Li, T. Pullerits, W. Deng, K. Han, *Angew. Chem. Int. Ed. Engl.* **2017**, 56, 12471.
[14] Y. Zhang, J. Yin, M. R. Parida, G. H. Ahmed, J. Pan, O. M. Bakr, J. L. Bredas, O. F. Mohammed, *J Phys Chem Lett* **2017**, 8, 3173.
[15] Z. Dang, J. Shamsi, F. Palazon, M. Imran, Q. A. Akkerman, S. Park, G. Bertoni, M. Prato, R. Brescia, L. Manna, *ACS Nano* **2017**, 11, 2124.
[16] T. Kawai, A. Ishii, T. Kitamura, S. Shimanuki, M. Iwata, Y. Ishibashi, *J. Phys. Soc. Jpn.* **1996**, 65, 1464.
[17] Z. Xiao, W. Meng, J. Wang, D. B. Mitzi, Y. Yan, *Materials Horizons* **2017**, 4, 206.
[18] A. R. Srimath Kandada, A. Petrozza, *Acc. Chem. Res.* **2016**, 49, 536.
[19] S. G. Motti, M. Gandini, A. J. Barker, J. M. Ball, A. R. Srimath Kandada, A. Petrozza, *ACS Energy Letters* **2016**, 1, 726.
[20] S. Öz, J.-C. Hebig, E. Jung, T. Singh, A. Lepcha, S. Olthof, F. Jan, Y. Gao, R. German, P. H. van Loosdrecht, *Sol. Energy Mater. Sol. Cells* **2016**, 158, 195.
[21] D. Meggiolaro, S. G. Motti, E. Mosconi, A. J. Barker, J. Ball, C. A. R. Perini, F. Deschler, A. Petrozza, F. De Angelis, *Energy & Environmental Science* **2018**, 11, 702.
[22] R. J. Mendelsberg, M. W. Allen, S. M. Durbin, R. J. Reeves, *PhRvB* **2011**, 83.
[23] P. Tandon, S. Raudenkolb, R. H. Neubert, W. Rettig, S. Wartewig, *Chem. Phys. Lipids* **2001**, 109, 37.
[24] E. Lifshitz, *J Phys Chem Lett* **2015**, 6, 4336.

*Supporting Information for*

# Broadband Defects Emission and Enhanced Ligand Raman Scattering in 0D $Cs_3Bi_2I_9$ Colloidal Nanocrystals

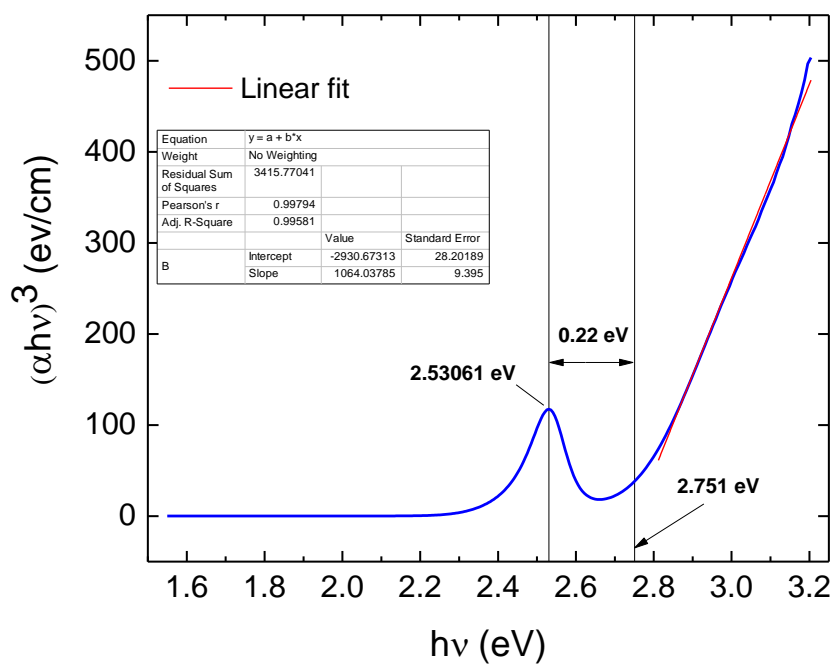

Figure S4. Tauc Plot for the NCs in toluene dispersion. The *r* value = 3 is consistent with an indirect band-gap.

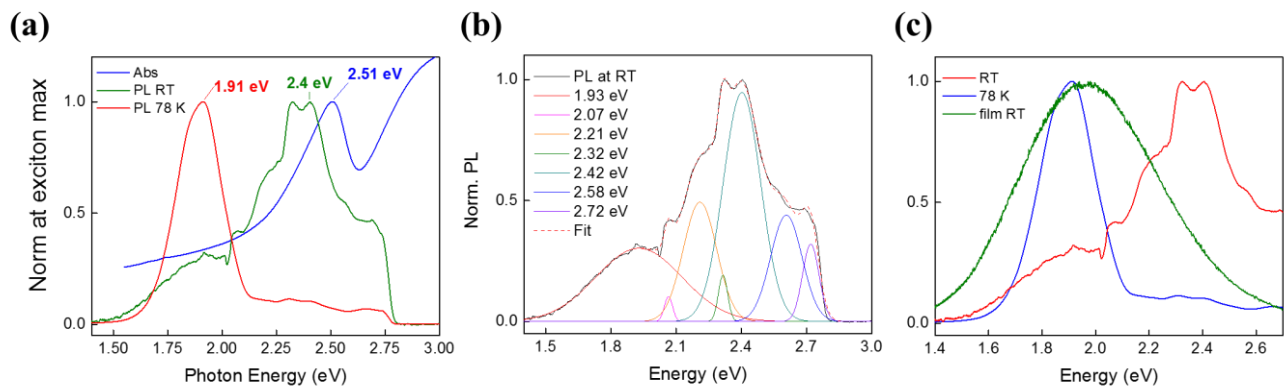

Figure S5. (a) Comparison between the UV-VIS absorption and the PL spectra at RT and 78 K (excitation 3.06 eV). (b) Deconvolution of the PL signal into seven Gaussian curves accounting for the oscillations incorporated in the emission band of $Cs_3Bi_2I_9$. Note that the last component (2.72 eV) can be linked to the Raman scattering of the oleate ligands, and it is poorly distinguishable due to re-absorption phenomena. For this reason, to decrease the re-absorption and gain more insights into this effect, we diluted the toluene dispersion 10 times (see main text). (c) PL of the NCs in toluene dispersion at RT (blue curve) and 78 K (red curve) and in solid film (green curve).

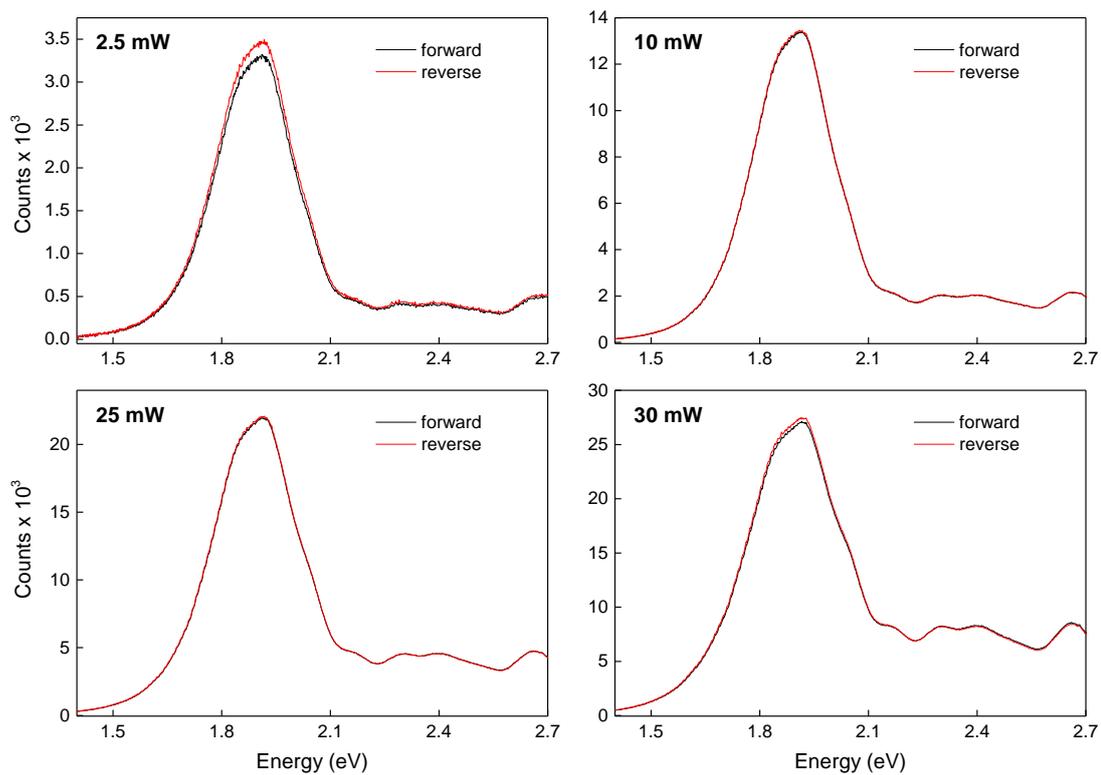

Figure S6. Comparison of the PL spectra at 78 K taken in forward and reverse scan, to investigate possible hysteretic behaviour in the emission of the NCs.

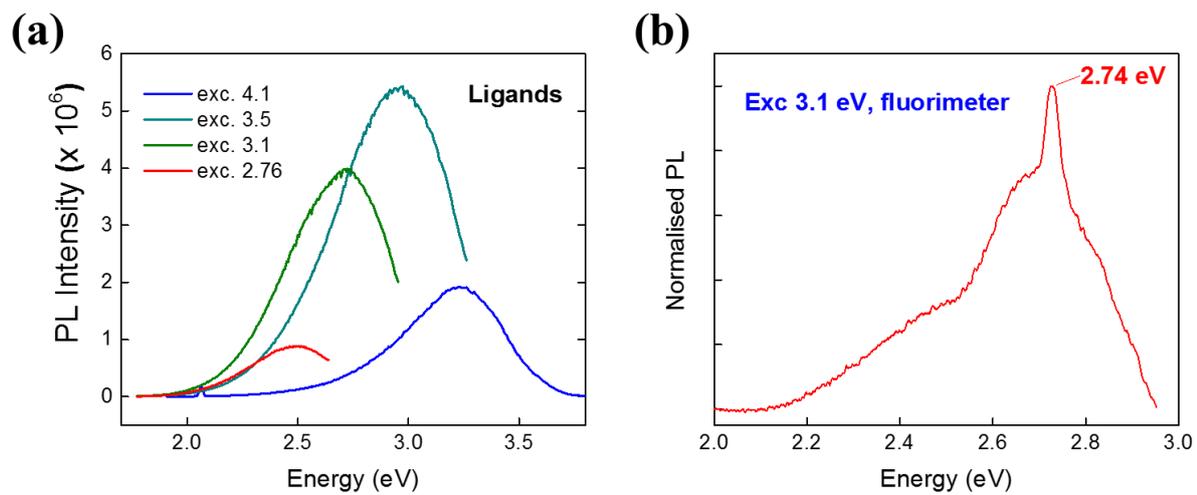

Figure S7. (a) PL of the oleate ligands only at various excitation energies. (b) PL of the NCs taken with the fluorimeter by exciting with a monochromatized Xenon lamp (25 μW).

19